# Understanding the Social Determinants of Mental Health of the Undergraduate Students in Bangladesh: Interview Study

Ananya Bhattacharjee, Department of Computer Science University of Toronto, Toronto, Canada

S M Taiabul Haque, School of Computer Science and Mathematics University of Central Missouri, Warrensburg, USA

Abdul Hady, Department of Computer Science and Engineering, Eastern University, Dhaka, Bangladesh

S.M. Raihanul Alam, Department of Computer Science and Engineering, Bangladesh University of Engineering and Technology, Dhaka, Bangladesh

Mashfiqui Rabbi, Health AI, Apple Inc, Seattle, USA

Muhammad Ashad Kabir, School of Computing, Mathematics and Engineering, Charles Sturt University, NSW, Australia

Syed Ishtiaque Ahmed, Department of Computer Science University of Toronto, Toronto, Canada

## Abstract

**Background:** Undergraduate student population is actively studied in digital mental health research. However, the existing literature primarily focused on students from developed nations, and undergraduates from developing nations remain understudied.

**Objective:** This study aims to identify the social determinants of mental health among undergraduate students in Bangladesh, a developing nation in South Asia. Our goal is to identify the broader social determinants of mental health among this population, study the manifestation of these determinants in their day-to-day life, and explore the feasibility of self-monitoring tools in helping them identify the specific factors or relationships that impact their mental health.

**Methods:** We conducted a 21-day study with 38 undergraduate students from seven universities in Bangladesh. We conducted two semi-structured interviews: one pre-study and one post-study. During the 21-day study, participants used an Android application to self-report and self-monitor their mood after each phone conversation. The app prompted participants to report their mood after each phone conversation and provided graphs and charts so that participants could independently review their mood and conversation patterns.

**Results:** Our results show that academics, family, job and economic condition, romantic relationship, and religion are the major social determinants of mental health among undergraduate students in Bangladesh. Our app helped the participants pinpoint the specific issues related to these factors as participants could review the pattern of their moods and emotions from past conversation history. Although our app does not provide any explicit recommendation, participants took certain steps on their own to improve their mental health (e.g., reduced the frequency of communication with certain persons).
**Conclusions:** While some of the factors (e.g., academics) were reported in prior studies conducted in the Global North, this paper sheds light on some new issues (e.g., extended family problems, religion, etc.) that are specific to the context of the Global South. Overall, the findings from this study would provide better insights for the researchers to design better solutions to help the younger population from this part of the world.

## Introduction

Like many low and middle-income countries (LMICs) [69], mental health among youth in Bangladesh is an overlooked and stigmatized topic [52]. More than 30% of adults in urban areas are struggling with mental health related issues [31]. At the same time, the mental health facilities and services in Bangladesh are also insufficient [23]. The National Institute of Mental Health (NIMH) and Pabna Mental Hospital are the only major institutes that provide mental health treatment, constituting only 700 beds in total [31]. There are some other mental health institutes in or around the big cities too, but two-third of the total population living in rural areas have difficulty accessing mental health due to lack of facilities in rural areas [31]. Support for non-serious mental illness is even poorer; the number of mental and emotional health support helplines is extremely low [36]. In addition, the general population still holds the traditional negative attitude and stigma towards mental health patients [3, 71].

Proposing a path forward for addressing mental health challenges in the Global South is still at a rudimentary stage of development [53]. The social determinants of mental health [16] – social and cultural factors which deeply impact one's mental health – have remained understudied in the context of Bangladesh. These factors vary depending on an individual's age, financial condition, or social surroundings. However, in this paper, we focus particularly on undergraduate students, who constitute a non-trivial part of the population and are susceptible to mental health related problems [62]. As various statistics and news reports corroborate the deep-rooted existence of suicide [59], depression [63], substance use [49], and extremism [22] among the young population in Bangladesh, and prior research indicates that such behavioral aberrations are often fueled by broader social, cultural, and political contexts in which one grows up and lives in [1, 5, 13, 37, 38, 46], it is timely to investigate these underlying factors.

Concerns over the mental health of the university students have been expressed in the global north for a while now [51], although the same cannot be said about the global south. According to a study conducted in a European country, around one-third of the first-year university students have been found to have mental health related problems [14]. Several studies [14, 19, 41] have associated academic performance with mental health, i.e. poor academic performance worsens mental health and vice versa. Other contributing factors for mental health deterioration include economic problems, high parental expectation, strained relationships, poor lifestyle, and so on [6, 11, 47, 61, 71]. Several mental disorders are frequently observed among the people aged between 14 and 24 years [39], many of whom are also university students. These students tend to encounter new experiences like moving away from home, making adult financial decisions, and so on. Besides, many of them experience changes in their health behaviors [70]. Adapting to these changes sometimes becomes difficult, as students are also expected to perform their coursework and participate in exams. Failure to cope with any of these challenges may cause mental health difficulties [43].

These factors have already been identified as the social determinants of mental health of students [16, 70], although they are often specific to the contexts of the global north. However, the different structure of the society and culture in the global south [7] may contribute to people's mental health in other ways that are specific to the contexts of the global south and not reported in previous studies. For example, in countries like Bangladesh, extended families are far more prevalent where people live with their relatives (i.e., uncles, aunts, cousins, or in-laws) [7]. Even when people are living in a nuclear family, the social communication among relatives or neighbors is more frequent and cordial. As a result, maintaining regular communication with even distant relatives is viewed as an important social duty [20]. Additionally, religion is a major component of life of people here [20]. The concepts of religion or religious duties differ significantly from Western context. A large portion of the population finds happiness through practicing religious duties, as religion offers their life a sense of value and purpose. Religious identity dictates an individual's social groups and their lifestyle choice [42]. The effects of complicated social relationships and disparate viewpoints on religion are yet to be understood in the context of student' mental health in the global south.

Through this work, we intend to explore the social determinants of mental health, particularly in the context of Bangladesh. In a country like Bangladesh where social and family structures are hierarchical, traditional, and community-based [16, 25, 70], social and family relationships play an important role in people's daily life, and recording one's mental condition after interactions with family members and social peers seems to be a useful way to monitor and improve their mental health. Past research suggests that patterns of mobile phone calls can reveal important information about an individual's mental state and relationship with their friends and family [27]. Besides, mobile phone communication history helps oneself identify various forms of social closeness [24]. In recent years, Bangladesh has observed a tremendous growth in mobile phone subscription. The total number of mobile phone subscribers in Bangladesh is around 160 million [10], which is almost equal to the total population of the country [9]. This shows that mobile phones are culturally relevant in Bangladesh. We see this as a research opportunity, as we can leverage the widespread use of mobile phones across the country for identifying the social determinants of mental health of undergraduate students in Bangladesh.

One potential way towards forwarding the research on identifying the factors that impact mental health can be the use of reflective tools [2, 17, 50] that support self-monitoring of mood and emotions [18]. These mobile applications are suitable for measuring subtle emotions and capturing patterns of behavior in long-term [18]. [39] conducted a study on people between the age of 14 and 24 and reported that self-monitoring one's mood results in an increase of emotional self-awareness (ESA), eventually reducing depression. Another similar study [40], which helped participants identify the cause of their drinking habit (e.g., relation with romantic partners), showed success in tracking information about alcohol consumption and its effect on mental behavior. [41] explored the impact of mood-tracking applications in clinical settings and their pilot study received positive feedback from both participants and therapists, as participants could review their relation with their family and peers. [54] suggested that reflective tools with the option to provide self-reported emotions can give good insights into human behavior. In psychology literature, Ecological Momentary Assessment (EMA) [60] supports the existence of a theoretical framework that demonstrates the importance of such in-the-moment mood assessment. The longitudinal nature of data in these tools gives users an opportunity to examine the effect of different life situations on their mental state, as users can document their mood and emotions at various times and do independent research on their historical data later [60]. [54] claims that providing information about one's emotion and mental state at random times of the day can lead to improvement in social behavior. Other studies [8, 15, 33, 57] also show the promise of self-reported mood tracking applications in mental health research.

In this work, we design, develop, and deploy a mood tracking Android application that could be used as a reflective tool to record and reflect on their past interactions with their social peers. We have conducted two semi-structured interviews with our participants - one before and one another after using the app to understand the impact of our app on their mental health. Our primary goal was to progress the research towards identifying the factors that were affecting the mental health of Bangladeshi undergraduate students. However, we do not make any causal claims about these factors, and instead we present the findings from our interviews that give insights into how certain factors manifest in the specific context of Bangladesh as well as shed light into some new issues that were rarely discussed before. As a secondary goal of the study, we wanted to observe the effect of the self-monitoring tool and see whether our participants take any action to improve their mental health, despite no explicit recommendation provided by the application.

## Methods

### Participants

The study was conducted on 38 participants between the age of 19 and 27. At the time of the study, all the participants were full-time undergraduate students. Participants were recruited through snowball sampling [26]– via social acquaintances, classroom announcements, and word-of-mouth. Students were recruited from seven universities at two major cities in Bangladesh: Bangladesh University of Engineering and Technology (BUET), Eastern University (EU), TMSS Medical College (TMC), Dhaka College (DC), Rajshahi University (RU), City College (CC), and Bangladesh University of Business and Technology (BUBT).

Our participants included 30 traditional students and 8 non-traditional students (P1, P5, P24, P25, P26, P30, P35). In Bangladesh, traditional students join university immediately after receiving their high school diploma, whereas most non-traditional students attend university after receiving two to three years of technical vocational training (e.g., Computer Office Application) and a few years of job experience [35]. Table 1 provides information about our participants' gender, age, and institution.

Table 1. Information about the participants. In the column 'Student Type', 'N' denotes non-traditional student and 'T' denotes traditional student. P36-P38 did not take part in exit interviews.

| Participant | Gender | Age | Institution | Student Type |
|---|---|---|---|---|
| P1 | Male | 25 | EU | N |
| P2 | Male | 24 | EU | T |
| P3 | Male | 22 | EU | T |
| P4 | Female | 24 | EU | T |
| P5 | Male | 25 | EU | N |
| P6 | Female | 24 | EU | T |
| P7 | Female | 25 | EU | T |
| P8 | Male | 23 | BUET | T |
| P9 | Male | 23 | BUET | T |
| P10 | Male | 24 | BUET | T |
| P11 | Male | 23 | BUET | T |
| P12 | Male | 23 | BUET | T |
| P13 | Male | 23 | BUET | T |
| P14 | Male | 24 | BUET | T |
| P15 | Female | 23 | BUET | T |
| P16 | Female | 23 | BUET | T |
| P17 | Female | 23 | BUET | T |
| P18 | Female | 20 | BUET | T |
| P19 | Male | 23 | BUET | T |

| P20 | Female | 22 | RU | T |
| P21 | Male | 22 | DC | T |
| P22 | Male | 21 | DC | T |
| P23 | Male | 22 | DC | T |
| P24 | Female | 24 | EU | N |
| P25 | Male | 27 | EU | N |
| P26 | Male | 20 | EU | N |
| P27 | Female | 22 | CC | T |
| P28 | Male | 19 | RUET | T |
| P29 | Male | 20 | RUET | T |
| P30 | Male | 25 | BUBT | N |
| P31 | Female | 19 | TMC | T |
| P32 | Female | 19 | TMC | T |
| P33 | Female | 19 | TMC | T |
| P34 | Female | 19 | TMC | T |
| P35 | Male | 26 | EU | N |
| P36 | Female | 23 | EU | T |
| P37 | Male | 23 | BUET | T |
| P38 | Female | 20 | TMSSDC | T |

## Study Design

We designed and developed "HWC" (How Was the Call) – a mobile phone application which is a self-reporting tool to record and reflect on past phone calls. We conducted two semi-structured interviews with each participant, one before (baseline interview) and another after (exit interview) using the app. Between these two interviews, the participant was asked to install and use the app for three weeks. The whole study was approved by the Institutional Review Board (IRB) of the authors' institution.

### Baseline Interview

During this interview, we asked participants about their general mental health and the reasons behind their mental stress, depression or frustration, the frequency of experiencing mental health issues, the relationships that cause them, their coping strategies, etc. As English is the primary medium of instruction in their institutions, all of our participants are proficient in English. However, to collect spontaneous responses and have an in-depth discussion, we conducted our interviews in their native language Bengali – the official language of Bangladesh [44]. We occasionally used English words for different mobile phone and mental health related terms.

Interviews were conducted both in person and through Zoom Video Conferencing Platform. All interviews were audio-recorded with the consent of the interviewees. Interviewers ensured appropriate levels of empathy and rapport during the interviews. Timeouts were offered if participants asked for them.

The average length of the interviews was 25 minutes.

### *Three-week study with HWC app*

After the baseline interview, participants were requested to use the HWC app for three weeks. Study personnel helped participants install the app on their smartphone and instructed them on how to use it. We designed the app to run in the background and pop up a survey at the end of each phone call. The survey asked participants to describe their mood where they could choose from eight emotions (Figure 1): *excited, cheerful, calm, neutral, bored, sad, tense, and irritated*. These eight emotions cover the circular space of valence-arousal dimension [56], a common framework for recording emotional experience, which assumes that all human emotions are distributed in a two-dimensional space. Prior works [18, 26, 54] show that this approach of asking emotions from valence-arousal dimension is particularly suitable for applications where users need to give frequent quick inputs on their mood and offers a reliable tool for researchers to analyze user behavior. One might argue that self-reported moods might not cover all mood archetypes, but a vast amount of psychology literature [18, 27, 54] validates the use of our or similar to our self-report methods. Other than the ease and convenience of the implementation of these methods, self-reported methods provide users with enough control over their interaction with the system, which does not make them feel they are part of an experiment and instead generate closer to real world data [18, 30]. These approaches inspire spontaneous responses from people and are helpful for capturing fluctuation of behavior and symptoms [30].

In addition to the eight emotions, we also provided participants with a text box to describe their feelings in more detail. While we could have asked for more inputs from users (e.g., how strong is the emotion?), our app asked for user input every time they make or receive a call. Hence, asking for too many inputs from users after each phone call would require high cognitive burden that might alter emotion and reduce long term engagement [18]. However, a participant can choose to skip and provide no information. This application does not require any internet connection and stores all information in the local memory of the mobile phone. As a result, there is no risk of participants' information going public.

Once the participants describe their feelings, they have the option to view their data later. Participants can review their call history and see how they felt after each call (Fig. 2). They can also view a visual summary of their reactions over a period of time with the help of a pie chart (Figure 2). Although the app lets participants see their information, it does not provide any sort of feedback or recommendation to alter mood.

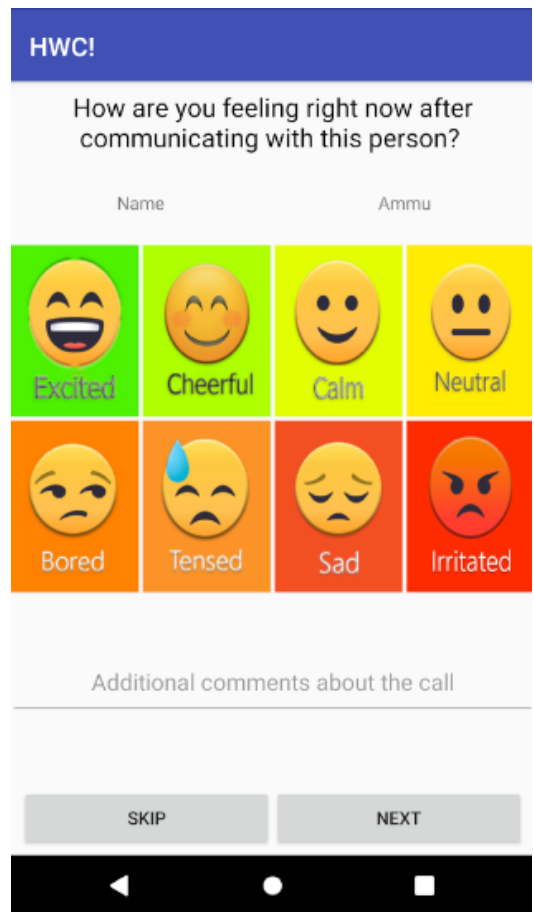

*Figure 1. Interface for providing feedback after each call*

The flow of the app is as follows:
- Participant installs the app.
- Every time the participant makes/receives a regular phone call, the app starts its operation.
- After the call ends, the participant is prompted to provide feedback and choose from eight options (Figure 1). A text box is provided for additional written comments as well. A participant can take any of the following two steps:
    - After providing the information, the participant can press "Next". This will store the information in the local memory of the mobile phone and the application will terminate.
    - Participant can choose to not provide any information by pressing "Skip". The application will terminate.
- Participant checks the call history and reviews his/her emotional status after each conversation at any time(Figure 2).

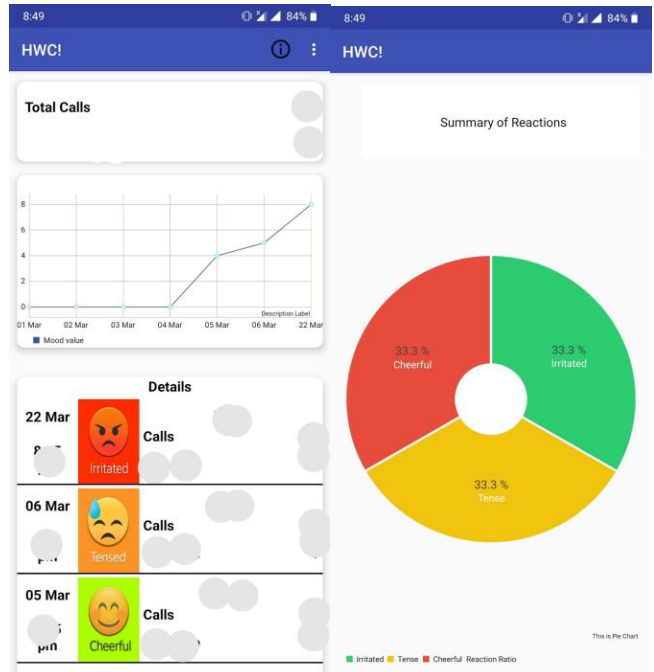
*Figure 2. Interfaces for analyzing call history and mood*

Participants, who were worried about their data being stored outside the phone (public server, database, etc.) or shared with a third party, were ensured that their call history and mood descriptions would only be stored in their mobile phone memory and even allowed to see the source code of the application. In fact, two participants who were familiar with android application development requested to review our source code and we allowed them to do so. We did not collect specific information about the number of calls received by each participant for privacy reasons as well.

### Exit Interview
Exit interviews were arranged after the participants had used the HWC app for three weeks. In this interview, participants described their experience of using the app to track their mood. The questions were more specific in this session where we asked them to highlight the major incidents that they had recorded over the past three weeks. We also asked them how the app had shaped their opinion and attitude towards other people in their social circle. The setting of this interview was the same as the baseline interview. The average length of the interviews was 17 minutes. Despite our best efforts, we could not reach three participants (P36-P38) in this session due to their unavailability, resulting in a data set of 35 (21 male, 14 female) interviewees.

### Data Analysis
We first transcribed the audio recordings from both sets of interviews and then translated those into English. Next, we performed thematic analysis on our data based on Boyatzis's framework for code development [12]. From our analysis of baseline interviews, we tried to explore the major social determinants of mental health among the undergraduate students in Bangladesh. The responses of the participants during the exit interview demonstrated the effectiveness of the app in monitoring their mental health and its impact on identifying the specific personal issues they had been dealing with. We describe the results of our analysis in the next section.

## Results

In this section, we first describe the findings of our baseline interview by highlighting the factors that impact the mental health of our participants. Then we report the impact of our app on helping participants monitor their mental health. Finally, we describe the measures that were taken by our participants to improve their mental health, although no recommendation was provided by the app.

### Factors Affecting Mental Health

Through our interviews, we identified five factors that impacted our participants' mental health the most. Figure 3 illustrates the distribution of these factors based on how many participants reported them. Several participants reported multiple factors. Below each of these factors is presented in detail.

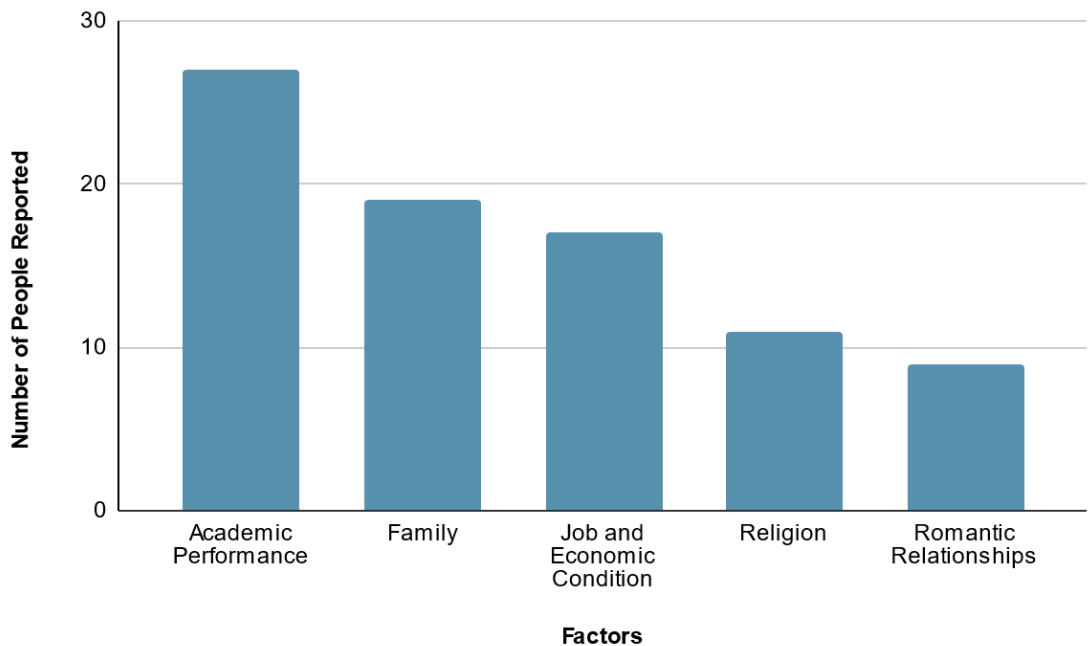

*Figure 3. Distribution of the factors that were reported by participants*

#### *Academics*

Academic standing has been the most discussed theme during the interviews. Overall, 27 out of 35 participants noted that academic performance has a major impact on their mental health. Their main concern was the stress and depression

created by assignments, class tests, projects, and final exams. One participant (P4) said:

*I feel depressed on the evening before my exams. I get very tensed and start to think too much about the upcoming exam. Often, I feel uncertain whether I will be able to answer all the questions or not. Sometimes I am too confused and start to panic about my preparation. I call my classmates and ask them about their preparation. Although they tell me that they have not prepared very well either, I assume otherwise and get more depressed.*

Another participant (P11) further added that he had been stressed because of his final year academic thesis. Even during vacation, he had to work for his thesis. Other participants were more concerned about their academic grades. They think that their results were not satisfactory, and they might struggle in the job market. As P2 mentioned:

*I get frustrated when I try to read in the evening. I think about my future career and feel that I am not doing enough. I have a poor CGPA, will anyone give me a job? At times, it seems that I am a complete failure, which makes me frustrated.*

A similar sentiment was reflected in the response of P10. In his opinion, despite putting his best efforts, he had not been able to achieve his desired grades, which resulted in mental depression and frustration. He said:

*I have been getting poor grades in exams in consecutive semesters. Despite trying my best, I am not being able to improve my grades. It's not that I'm not trying, but nothing seems to work.*

Another participant (P17) expressed her concerns over academic deadlines. She feels that procrastination contributes to the decline of her mental health. She tends to start working at the eleventh hour to complete her assignments but before that she cannot focus on anything else either as she constantly keeps reminding herself about the deadline. She feels that this sort of behavior has put her in "some kind of loop", which is the main reason behind her stress. P25 also shared similar frustrations, but his reason was he did not have time to focus on his studies because of his full-time job. He also felt that he was in a loop because he joined the program to have a BSc degree and have promotions in his current career, but at the same time his career prevents him from concentrating on his studies.

### *Family*

Nineteen participants mentioned family as a reason behind the decline of their mental health. However, the underlying reasons varied significantly. Since the participants were from different backgrounds in terms of social and economic status, their problems were also different in nature. For example, one participant (P1) said that he had lost his parents at the age of 15. As the eldest child, he had to pick up many responsibilities since then. Although he is quite happy to perform his responsibilities, he constantly feels that he has not been doing enough for a better upbringing of his younger brother. He said:

*When my mother died, my younger brother was still a baby. We lost our father after a few years, so his only guardian now is me. I am always tense about his education. Sometimes my brother does not listen to what I say. Those times are very hard for me. Whenever he fails at anything, I feel genuinely frustrated.*

The participants who were married (P5, P6, P7, and P24) talked about their spouse and in-laws when discussing mental issues. One male student (P5) with a full-time job said that he is always concerned about financial stability, as his wife does not have a job. Due to his busy life, he misses many family events and although the reasons are genuine, his in-laws feel that he does not give them enough attention. Three recently married female participants (P6, P7, and P24) discussed at length about their relationship with in-laws and how this affects their mental health. In Bangladesh, it is a common practice for women to move to the husbands' house after the wedding. As extended families are still a cultural norm in Bangladesh [7], a new wife needs to build relationship with her husband's family. These issues were pointed out by both participants. As P7 said:

*I am currently having issues with my new family. It is not any particular person, but their overall attitude. I have to be careful about my every action because if they don't like anything, they directly call my parents to complain about me. What hurts me more is that my own parents do not support me in this case. The relationship dynamic with my in-laws is perhaps the biggest cause of stress in my life currently.*

Some other participants mentioned that as they are too busy with their academic life, they get stressed whenever they are burdened with any additional family responsibilities, including sending money and taking care of sick family members. Two participants (P13 and P27) said that simply talking about family problems through mobile phone is stressful for them.

### *Job and Economic Condition*

Five of our participants (P1, P5, P25, P30, P35; all non-traditional students) had full-time jobs during the study period, and all remarked that work-related issues affect their mental health. P5 talked about the competitive environment of his office – how everyone there is under huge pressure to perform and complete the tasks assigned to them. He said:

*A full-time employee has a lot of responsibilities. You have pressure from your boss – pressure of meeting deadlines, pressure of performing better than your colleagues. Even at midnight, I receive calls from my boss. This makes me think that life without mobile phones would have been much better. At least your boss could not have reached you after office hours. … When I fail to meet a deadline, I get anxious. You fail, which means that you will have a lesser chance of being promoted and there are many other*

*people waiting to grab that opportunity. I used to think competition exists only in colleges, but it is certainly more in professional life.*

Some of our participants mentioned that their families are solvent, and they do not have an urgency to get internships or jobs. However, they still feel the pressure to look for a job as their peers have been doing the same. One participant (P14) highlighted this issue as a major reason behind the decline of his mental health. Some participants, on the other hand, get confused about their career choices. In Bangladesh, career counseling is not recognized as an important activity at universities [4, 32], and thus, many students face a dilemma regarding their career choices. This has been reflected in the response of a business major (P6):

*I think about my future all the time. I am finding it really difficult to choose between two specific options: doing an MBA or preparing for job interviews. I try to consult my seniors, but even they do not have a clear career path for this program.*

As mentioned before, thirteen of our participants are from BUET, which is typically regarded as the most sought-after academic institution in the country. As a result, they have a high demand as instructors in local coaching centers [67] or as private tutors. In fact, all of our participants from BUET except P9 had a part-time income source as a private tutor. We found a few tutoring related issues that affect their mental health. For example, P10 mentioned that he receives a lot of calls from his students during his own study hours and concentrating too much on his students might be a reason for the deterioration of his mental health as it affects his own grades. P11 also expressed similar concerns but for him, making money is a bigger priority than obtaining good grades, even as a student. In his opinion, he becomes frustrated when he does not have a good number of students to be able to send enough money to his family. In fact, he was mentally quite happy during our study period as he had been earning more money than ever.

### Religion

Eleven participants (all Muslims) discussed the role of religion in their mental health. One of them (P1) said that when he misses the Fajr prayer (the first of the five daily prayers performed by a practicing Muslim) at dawn for oversleeping, he does not feel well for the rest of the day. However, he eventually calms himself by thinking that an unwillingly committed negligence would not be punished and a good act later in the day would compensate for the earlier remission. He further added:

*I believe in Tawhid* (the concept of monotheism in Islam)*, so I believe the whole universe has been created by our God. He has given us certain duties to perform, and sometimes when I fail to do any of those, I get a little bit stressed out. ... I have another issue pertinent to religion. In a society, different types of religions coexist, and I have certain expectations from the believers of other religions. At times, those expectations*

*are not met, and I face trouble in interacting with people from other religions, which bothers me a lot mentally.*

P27, who also expressed similar sentiment, felt that each problem in life cannot be shared with family. That is why he had turned to God with Whom he feels every problem can be shared and discussed. However, all of these participants highlighted that they perform Salah (prayer) to seek happiness. Additionally, P4 and P24 said that they regularly read the Quran (the main religious text of Islam) to relieve mental stress.

### *Romantic Relationships*

Nine of our participants talked about their romantic relationship when discussing mental health. The three married participants, including P7 who was not happy with her in-laws, appreciated the support that they received from their spouse in conjugal life. However, all of them confessed that they need to think about their actions thoroughly as those might affect their spouse. This constant pressure of remaining careful and alert creates a stressful family environment for them.

The unmarried participants (P9, P11, P13, P21, P28, and P31) described their past and present romantic relationships. One participant (P21) correlated his stress and frustration with the relationship with his partner saying,

*"I think my major source of stress and depression is my relationship with my girlfriend. We have been in the relationship for a while now, and I always try to make her happy in any way I can. Even then, I sometimes notice that the value of my opinion matters less very little in our relationship. Additionally, I have a tendency to compare my financial status with my girlfriend's, which also makes me sad or in some cases, jealous."*

Another participant (P9) pointed out a different aspect of relationships saying that his ex-girlfriend had issues of emotional dependency and he had to spend a significant amount of time throughout the day talking with her over the phone. He felt relieved to get out of the relationship because he was unable to cope with the habit of spending too much time on the phone. P11, who had recently broken up with his girlfriend, was visibly distressed because of the failed relationship. However, one participant (P13) attributed the supportive nature of his girlfriend to his sound mental health saying that he always found his partner by his side when he was stressed.

### Monitoring Mental Health

Having described the factors affecting the mental health of our participants, we now highlight the effectiveness of our app in monitoring their mental health. Through our app, participants were able to track their own behavioral patterns by regularly recording their information. All of our participants agreed that the moods shown in the application represented the entire spectrum of their real-life emotions and made them more conscious about their mental health. As P1 said:

*Each time I finish a call, the app asks me how I am feeling. This makes me think about my mental health for a few moments, which I would not have done previously. The app makes me more aware of my mental health.*

Another participant (P2) liked the feature of reviewing the mood data. He reported that during the course of three weeks, he had periodically analyzed his mood report and in doing so, he could identify some patterns. For example, he had realized that he feels more tense at night.

Our participants also mentioned that the app helped them in identifying the positive and negative influences in their life. As a result of using the app, they were able to more consciously identify the positive influencers – persons with whom they are close in real life but never previously thought of as positive catalysts for the betterment of their mental health. For example, P5 acknowledged this revelation in the following way:

*Through this app, I could identify those people who make me happy and I realized that even during face-to-face conversations, this same group of people makes me feel more comfortable. I could make this explicit connection by using this app.*

Similarly, some participants noted that the use of the app had been helpful in finding out the persons who cause mental stress in their life. Two participants (P6 and P26) mentioned that the emojis were quite helpful in this regard as they could properly categorize their mental state by using those emojis. They were able to identify the conversations that had a negative impact by looking at the emojis for "sad" and "irritated" reactions. A couple of other participants (P2 and P4) also explicitly recognized the importance of the emojis. Choosing emojis after each phone call gave P4 a sense of excitement similar to "playing games on a mobile phone", which she viewed positively.

Some of our participants reported how the app assisted them in precisely identifying the reason for their mental stress. For example, during the baseline interview, P11 said that the final-year thesis had been his primary academic concern. By using the app, he was able to understand that his supervisor had been pushing him a lot for the thesis. After each phone call with his supervisor, he was either "tense" or "sad". At times, he felt "irritated" after receiving several calls regarding the improvement of the thesis document. Similarly, before using the app, P13 already knew that talking with his family members over the phone had been stressful for him. After using the app for three weeks, he realized that things get worse when economic problems are discussed during any conversation, but other times the conversations are not as stressful. Seven other participants also (P3, P10, P18, P22, P28, P30, and P35) specifically mentioned that the app helped them pinpoint the exact cause of their mental stress.

One participant (P12) provided an important feedback regarding the text box feature for writing additional comments. He thinks that the overall mood is not entirely dependent on the conversation as it gets affected by other factors (weather or an upcoming exam, for example) as well. He particularly liked the option of having a text box to be able to record the potential reasons behind certain emotions. In his opinion, the written comments were helpful to better understand the contexts when he reviewed those later.

### Impact on Mental Health Improvement

As the use of the app enabled our participants to better identify the persons or events that had a negative impact on their mental health, they could also take some measures for improvement. About two-thirds (24 out of 35) of our participants said that they had tried to change the frequency of contacting those people who were having a negative effect on their mental health. As P6 said:

*The app helped me identify the people who were causing stress. As I checked the call patterns, I tried to understand why I was not feeling comfortable talking with them. What I did was – I tried to improve my relationship with them. Some attempts succeeded, while others did not. I have started avoiding those persons with whom I could not develop a better relationship.*

The responses from P14 also revealed a similar adjustment. By using the app, he realized that he had been in touch with some friends who were detrimental to his mental health as phone conversations with them were making him "sad". He reduced the frequency of phone conversations with those people and reported that he had already started feeling better. P29 was also able to identify such people in his life. Apart from having less phone conversations with them, he started changing his tone and approach in physical meetings. He said,

*After I found out who those friends were, whenever I met them, I explicitly tried to show that I am not interested in talking with them. I deliberately made the conversations shorter and removed the friendly approach from my tone.*

However, our participants also mentioned that it is impossible to avoid certain persons in academic or professional lives (supervisors and managers, for example), and they had to pick up phone calls from those persons or call them for important academic or professional reasons. At times, they also had to pick up specific calls for courtesy and modesty. Some participants mentioned that the app had been helpful in these cases as based on prior mood patterns recorded through the app, they were mentally prepared for certain conversations to have a potential negative impact on their mental condition. In words of P5,

*In my office, I often need to contact many colleagues and those conversations can't be avoided. But now the app has let me know that after some of those conversations, I'll not feel cheerful. As I know this beforehand, the impact is less severe.*

One participant (P1) said that the app helped him control his temper. He talked about a particular relative who used to call at inappropriate times for mundane conversations. As P1 felt "annoyed" after receiving that relative's call, he tried his best to remain calm when talking with him. P1 also mentioned that the app made him aware of his attitude towards a few other family members as well. Another participant (P10) said that the app offered him the opportunity to find a new support system within his own family. He said:

*I have a large family with many sisters. As I am under huge academic pressure, I cannot contact them as frequently as I would like to. But after using the app, I found that my stress relieves a lot when I share my problems with my sisters. They give wonderful advice, too. I have decided to communicate more frequently with them.*

The app also seemed to have a positive impact on the relationship status of a participant (P13). Although P13 had a stable and healthy relationship and used to share his problems with his girlfriend even before using the app, he noted that the app reinforced the same notion about the relationship, as he often felt "cheerful" or "calm" after talking with her.

## Discussion

### Principal Results

#### *Social Determinants of Mental Health of the Undergraduate Students in Bangladesh*
Our baseline interviews identified several factors that affect the mental health of the students in Bangladesh. Our findings demonstrate some differences between Western countries and a Global South country like Bangladesh in regard to social determinants of mental health of the young, college-going population. While some of the factors (e.g., academic performance [14, 19, 21, 41] and economic condition [47, 66]) we identified in our study have already been pointed out by prior studies conducted in Western contexts, issues such as extended family (in-laws) problems and religion have not been reported before.

Female students, who go for higher education, need to overcome the barriers created by the local society [34, 64]. The situation is even more difficult for married students. In a country, where more than 52% children get married before reaching 18 years old [45, 68], it is expected that a large number of undergraduate students, especially the females, would be married. We had three married students in our study, two of whom are female. According to the local culture of Bangladesh, the bride moves in with the groom's family after marriage. New brides often need to adjust in a different family and adapt to their culture and rules. When discussing mental health, both of our female married participants talked about their struggle to adapt themselves to their husbands' family. This struggle to adapt with in-laws or extended families should be investigated in the context of students from other countries in nearby regions (e.g., India, Pakistan), as underage marriage and discrimination against women likely exist in those countries too [58, 69].

Religion has been an integral part of the social, political, and economic environment of Bangladesh [20]. The state religion of the country is Islam [65], and 90.3% of the total population identify themselves as Muslim [9]. Religious identity is a significant factor behind happiness in Bangladesh [20]. This view aligns with one of our participants' stress behind failure to perform religious duties. Some other participants also briefly mentioned that they perform regular prayers to seek mental happiness.

Although economic condition has been a factor discussed in previous studies, the context of Bangladesh is somewhat different. In the local society, a male is expected to support his parents and, in some cases, the extended family. So, students at the undergraduate level, especially the male ones, start worrying about their financial status. Families also start sharing financial problems with them, indicating that the students now need to contribute to their family. For example, P11 and P13 felt stressed and frustrated when they failed to send enough money to their home. However, we did not get such reports of stress over family's economic status from any of our female participants.

Apart from traditional students, we also had eight non-traditional students (One female and seven males) in our study. Like regular students, they brought up issues about course works and exams, but all of them pointed out other issues as well. Four of them were married, so these participants mentioned their struggle to measure up to their in-laws. Even the male married student (P5) talked about how his inability to attend family functions makes his in-laws unhappy. Five male participants (P1, P5, P25, P30, and P35) explained their job life in detail, and expressed that they find it very difficult to manage a full-time job with academic life. Their conversations (particularly P25's) hinted that they were expected to do a full-time job as they were getting old, but at the same time they had to be serious about their academic coursework, as not having the bachelor degree hindered their progress in job life.

### *How does the App Help Identify One's Own Social Determinants?*
The most commonly appreciated theme regarding our application was its ability to help users observe their mental states. In the baseline interviews, many of our participants had rough ideas about the main factors affecting their mental health, but using the app for three weeks helped them pinpoint the exact reason. During this time period, our participants had been constantly putting their emotions after each call. As our app provided visual analysis of call history through graphs and pie-charts, they could interpret those in their own way. Our participants were able to notice which calls evoke negative sentiments and act accordingly. For example, the app helped several participants identify that academic supervisor or some specific family members are the reason of their stress. Conversely, the app also helped them identify the relationships that have a positive impact on their mental health. One participant further discovered the temporal aspect of mental health by using our app.

Similar to other reflective tools [2, 17, 50] with the option to provide self-reported emotions, one benefit of our application was that participants could analyze their calls and moods in any way they wanted. Looking at the graphs and charts, they could review their moods over multiple days, or they had the option to analyze their conversation with a single individual. Although in our study participants mostly talked about identifying positive or negative relationships, we also had people who identified other patterns of their mental states (e.g., P2 felt more stressed at night). P13 identified that talking with family only stresses him when the conversation is about financial problems. Overall, the flexibility that we provided our participants regarding interpreting their calls and emotions enabled us to have a diverse set of insights into their mental state.

Although the app was used only by Bangladeshi students in our study, apps like these can be used in a universal context. In our study, students could identify the factors causing negative impact on their mental health. These factors may be different in the context of people from other cultures, but everyone can realize their own factors by using the app for a prolonged period of time. Even in our study, the participants reported a range of factors using the same application. We believe that this app can be considered as a variant of digital diary [48], which can act as a support system for mental health [62]. However, we note that the topic of using the textbox to describe the phone calls has come up only one time in our study.

### *Actions Taken by Users*

After identifying some of the potential factors, participants in our study took some measures to improve their mental health. These measures were not suggested or recommended by our app, rather those were taken by the participants on their own. For example, 24 students reported that they changed the frequency of contacting the people that they found to have a negative impact in their life. Among these participants, some tried adjusting their stressful relationships, while others started avoiding interactions altogether. In the cases when avoiding a person had not been a feasible option because of professional reasons (for example, colleagues or supervisors), the app helped the participants prepare themselves mentally for a potential negative conversation.

It should also be noted that the app cannot address all cultural or social aspects of human life. For example, our mobile application cannot directly assist the female students regarding the challenges they face in their day-to-day life. Treating women as inferior or violence against them are deeply rooted social problems and no mobile application can solve these issues immediately [64]. While our app can provide some information about the underlying reasons in contexts such as this, it cannot go beyond to fix those problems.

## Limitations and Future Work

Our paper does not make any causal claims about the social determinants of mental health (e.g., we do not make claims like marriage causes mental frustration to young women). We only describe the findings of our interviews and the factors we reported may not be causally related to mental health. Other hidden unmeasured variables might play a role. However, we believe our reported factors will pave ways for future work which will dive deeply into investigating underlying mechanisms of mental health, particularly the ones that came up for the first time in this study. We believe follow-up work can then inform the theories of mechanism of mental health. These theories can then be verified with actual field experiments to quantitatively find causal relationships.

Although we included students from seven universities that offer different standards of education, future works should interact with students from even more diverse backgrounds. Our participants predominantly live in two major cities, and students from universities in semi-urban or rural areas might face other factors that have not been identified in this study. It would also be interesting to observe how this different demographic group struggles with the factors reported in this study, and whether their perceptions about using a mobile application to monitor mental health would be any different.

Finally, there is a growing tendency among students to use social media platforms like Facebook Messenger, WhatsApp, or Emo for having longer conversations. As the Internet has started reaching even the most remote parts of the country [29], data plans and broadband Internet are not as costly as those were before and calling over Internet is a much cheaper option in many cases. Our participants also mentioned that they use several platforms to contact friends or family members. Future studies can be designed on these Internet platforms to monitor the mental health of the young, college-going population of Bangladesh in a more comprehensive way.

## Conclusions

In this work, we aim to progress the research towards identifying the social determinants of mental health of the undergraduate students in Bangladesh. We design and deploy an Android application among our participants to help them record and later reflect on the mobile phone conversations with their friends, family members, and academic/professional correspondences. Our app assisted the participants in pinpointing the exact relationship or factor that has a detrimental effect on their mental health. While some of these factors have been reported in the prior studies conducted in the Western contexts, we identify several new factors, including religion and extended family affairs, which are pertinent to the society of Bangladesh. Although our app does not provide any recommendations from its side, some participants took independent measures in order to improve their mental health. However, in some cases (e.g., extended family problems), despite identifying the problem, participants could not find an appropriate solution, but they could better prepare themselves to cope with that specific issue. Taken together, our study provides useful perspectives and insights on the mental health of the undergraduate students in Bangladesh, and we hope our findings can help researchers design better solutions to improve the mental health of the younger population from this part of the world.

### Acknowledgements
The project was supported by Discovery Grant, Natural Sciences & Engineering Research Council, Grant No. RGPIN-2018-0. The authors wish to thank Abdul Kawser Tushar (University of Toronto), Eftekher Ahmed (CUET), Mainul Islam (Charles Sturt University), and Md. Sabir Hossain (CUET) for their contributions in developing the app.

### Conflicts of Interest
None Declared.

### Abbreviations

BUBT: Bangladesh University of Business and Technology

BUET: Bangladesh University of Engineering and Technology

CC: City College

DC: Dhaka College

EU: Eastern University

HWC: How Was the Call?

JMIR: Journal of Medical Internet Research

RU: Rajshahi University

TMC: TMSS Medical College